# Dark matter cavity and amplification of gravitational waves

Xiao-qing Li

Department of Physics, Nanjing Normal University, Nanjing 210097, China

Abstract

The linearized Einstein theory in the approximately flat time-space is modified accounting for a massive graviton as a dark matter in the universe. On the basis of magnetic-type Maxwell- Proca -Vlasov equations, the interactions of the gravitoelectromagnetic fields (GEM), involving the vector potential, with gravitons (dark matter) dealing with resonant wave-particle and wave-wave one are studied. It is found that the behavior of the perturbed GEM fields and the perturbed matter density can be described by the nonlinear coupling Eqs.(4.6) and(4.7). It is shown that the perturbed field is instable with respect to the collapse, leading to enhancement of the GE field. In other words, the gravitational waves (GWs) are amplified and concurrently the cluster and cavity of the dark matter formed. And it is referred that the lower limit for graviton mass is $M_{gr} \geq 10^{-50}(g)$ by the condition to exist stable longitudinal mode.

    Modern cosmology based on the gravitational theory of Einstein, no doubt gets a great success. To explain observation phenomena in the large-scale, however, introducing exotic mass density, the so-called dark matter, is essential. One may consider consistently the gravitational field "quantum", the graviton, as a dark matter (Dubovsky et al., 2005). This may mean that in this case, need a modified gravitational theory with graviton of non-zero mass as a massive dark matter.

    Astrophysical observations indicate that dark matter constitutes most of the mass in our universe, but its nature remains unknown. Due to the huge masses, it has to play a very important role for evolution of celestial bodies and the universe. It is shown that the distributions of dark matter are apparent inhomogeneities; they have tended to cluster (Dubovsky et al., 2005). For example, there is an apparent minimum dark halo mass of ~ $2 \times 10^7 M_\odot$ ( Mateo *et al.*, 1993). And there is irregular density profile with surprisingly low and central density (Gilmore *et al.* 2007 ).

    Hence we should establish a model for large scale region far from the source, where Einstein's theory takes approximately relevant forms in consideration of the effects of the graviton with non-zero mass. Historically, de Broglie ( 1940) had proposed that there are photons with no-zero rest mass, $m_{pho} \neq 0$, which will result in important modification for electromagnetic Maxwell equation: adding the term ($\mu_\gamma^2 \phi$) and ($\mu_\gamma^2 \mathbf{A}$) in electric and magnetic field equations with source terms, respectively. This is the Proca equation (Proca, 1937). Similarly, in the case when

the gravitons possess non-zero rest mass, a really simple result analogous to the Proca equation is obtained (Argyris and Ciubotariu, 1997; Demir and Tanış, 2011).

Linearize Einstein field equations predicted the existence of gravitational waves with spin-2; the graviton, similar to the photon (the electro-magnetic quantum), propagates at a constant speed of light. The many efforts that have been made to detect gravitational waves have so far given no convincing evidence that they have actually been seen. In the late 1960s and early 1970s, Weber announced that he had recorded simultaneous oscillations in detectors 1000km apart, waves he believed originated from an astrophysical event. But many physicists were sceptical about the results that were three orders of magnitude higher than were theoretically predicted. In addition, observed high energies of gravitational waves by Weber have not been confirmed by these independent measurements. Since then, more than 30 years, many experimental physicists, more than 14 times, claim to have detected gravitational waves (Collins, 2004). This is perhaps, due to the fact that gravitational waves with high energies are very rare than many physicists had expected. On the other hand, it is highly possible that the gravitational waves can get largely increase the rate as the waves interact with interstellar matter, so that Weber results, perhaps, are acceptable.

Large scale gravitating system, such as has been speculated (Argyris and Ciubotariu, 1997), is a collection of dust particles and non-zero rest mass gravitons which exhibits collective mode behavior. In order to analyse the nonlinear interactions of many-particle collection on the basis of kinetics (Li, et al., 2009; Li and Liu, 2012), we need to derive the standard GEM - Proca equation. Research on non-linear interactions of medium and graviton with mass ($\mu_g = \frac{M_{gr} c}{\hbar}$) will may provide effective clues for gravitational waves detection and analysis.

## I. Gravito-Maxwell-Proca field equations

There are two kinds of approximate methods in gravitational theory: post Newtonian approximation and weak field one. Astrophysical accretion disks, far from the gravitational source (such as black holes and neutron stars), are suitable intersection region for the two approximations. In the case of a weak field and sub-light speed movement:

$$R \gg r_{sch} \equiv 2GM/c^2, \tag{1.1}$$

$$v \ll c, \tag{1.2}$$

the metric $g_{\mu\nu}$ can be expanded in powers of $\varepsilon = \overline{v}(\sim \sqrt{GM/\overline{r}})$,

$$g_{00} = -1 + \overset{(2)}{g}_{00} + \overset{(4)}{g}_{00} + O(\varepsilon^6), \quad g_{0i} = \overset{(3)}{g}_{0i} + O(\varepsilon^5), \quad g_{ij} = \delta_{ij} + \overset{(2)}{g}_{ij} + O(\varepsilon^4),$$

accurate to terms of $O(v^3/c^3)$, Einstein field equations

$$R_{\mu\nu} = -\frac{8\pi G}{c^4}\left(T_{\mu\nu} - \frac{1}{2}g_{\mu\nu}T^{\lambda}_{\ \lambda}\right)$$

are reduced (Weinberg, 1972) as

$$\nabla^2 \phi = 4\pi G \rho, \quad \nabla^2 \mathbf{A} = 16\pi G \rho \mathbf{v}/c \tag{1.3}$$

with

$$4\frac{1}{c}\frac{\partial \phi}{\partial t} + \nabla \cdot \mathbf{A} = 0, \tag{1.4}$$

$$\frac{d\mathbf{v}}{dt} = -\nabla \phi - \frac{1}{c}\frac{\partial \mathbf{A}}{\partial t} + \frac{\mathbf{v}}{c} \times (\nabla \times \mathbf{A}), \tag{1.5}$$

with the vector potential $A_i/c^2 \equiv \overset{(3)}{g}_{i0}$. The corresponding metric is

$$g_{\mu\nu} = \eta_{\mu\nu} + h_{\mu\nu} = \begin{pmatrix} -1 & 0 & 0 & 0 \\ 0 & 1 & 0 & 0 \\ 0 & 0 & 1 & 0 \\ 0 & 0 & 0 & 1 \end{pmatrix} + \begin{pmatrix} -2\phi/c^2 & A_1/c^2 & A_2/c^2 & A_3/c^2 \\ A_1/c^2 & -2\phi/c^2 & 0 & 0 \\ A_2/c^2 & 0 & -2\phi/c^2 & 0 \\ A_3/c^2 & 0 & 0 & -2\phi/c^2 \end{pmatrix}, \quad |h_{\mu\nu}| << |\eta_{\mu\nu}|. \tag{1.6}$$

Introducing electric and magnetic gravities (Braginsky *et al.*, 1977)

$$\mathbf{E}_g = -\nabla \phi - \frac{1}{c}\frac{\partial \mathbf{A}}{\partial t}, \quad \nabla \times \mathbf{A} = \mathbf{B}_g, \tag{1.7}$$

and using the harmonic coordinate condition(1.4), accurate to terms of $O(v^3/c^3)$, one gets

$$\nabla \cdot \mathbf{E}_g = 4\pi \rho_g, \quad \nabla \cdot \mathbf{B}_g = 0,$$

$$\nabla \times \mathbf{B}_g = \frac{16\pi}{c}\mathbf{j}_g + \frac{4}{c}\frac{\partial \mathbf{E}_g}{\partial t}, \quad \nabla \times \mathbf{E}_g = -\frac{1}{c}\frac{\partial}{\partial t}\mathbf{B}_g;$$

in which,

$$\rho_g = -G\rho, \quad \mathbf{j}_g = -G\rho\mathbf{v}.$$

This is GEM—Maxwell equations. Meanwhile the force $\mathbf{F}$ (1.5) on a unit mass, has Lorentz-type form

$$\frac{d\mathbf{v}}{dt} = \mathbf{E}_g + \frac{\mathbf{v}}{c} \times (\nabla \times \mathbf{B}_g). \tag{1.8}$$

Similar to the heavy--photon theory, considering the graviton with no-zero mass, we have

$$\nabla \cdot \mathbf{E}_g = 4\pi \rho_g - a\mu_g^2 \phi, \quad \nabla \cdot \mathbf{B}_g = 0, \tag{1.9}$$

$$\nabla \times \mathbf{B}_g = \frac{16\pi}{c}\mathbf{j}_g + \frac{4}{c}\frac{\partial \mathbf{E}_g}{\partial t} - \mathbf{Q} \quad , \quad \nabla \times \mathbf{E}_g = -\frac{1}{c}\frac{\partial}{\partial t}\mathbf{B}_g, \tag{1.10}$$

where,

$$M_{gr} \gg m_{gr}^0 \sim 10^{-52}(\text{g}); \tag{1.11}$$

by the use of continuity equation,

$$\frac{\partial}{\partial t}\rho_g + \nabla \cdot \mathbf{j}_g = 0 \tag{1.12}$$

one get from (1.9) and (1.10)

$$\mu_g^2 \frac{4}{c}\frac{\partial}{\partial t}\phi + \nabla \cdot \mathbf{Q} = 0;$$

via the harmonic coordinate condition (1.4),

$$\mathbf{Q} = \mu_g^2 \mathbf{A},$$

which is consistent with the Proca theory.

It must be pointed out that because the field equations (1.9) and (1.10) contain the vector potential and scalar potential, related to Lorentz gauge transformation ($f$ is arbitrary scale function),

$$\mathbf{A} \to \mathbf{A}' = \mathbf{A} + \nabla f, \quad \phi \to \phi' = \phi - \frac{1}{c}\frac{\partial}{\partial t}f,$$

hence it is not invariant: there can be no another degree of freedom for $\mathbf{A}$ and $\phi$.

Therefore the harmonic coordinate condition(1.4) limits the heavy gravitons with no-zero potentials($\mathbf{A}, \phi$), so that it is necessary to satisfy that $\nabla \cdot \mathbf{A} \neq 0$. This means the transverse gravitons have no contribution to the interaction fields ($\mathbf{E}_g, \mathbf{B}_g$) in the system.

Now one decomposes the total matter density and current in Eqs. (1.9) and(1.10):

$$\rho_g = -G\rho \equiv \tilde{\rho}_g + \rho_{g0}, \quad \mathbf{j}_g = -G\rho\mathbf{v} \equiv \tilde{\mathbf{j}}_g + \mathbf{j}_{g0};$$

in which the $\rho_{g0}$ and $\mathbf{j}_{g0}$ is the parts of an external field source in the medium, corresponding to an additional local mass disturbance, for example, by nonlinear interactions between the fields and the medium, or some test particles. In this case, it is convenient to define the vector quantity $\mathbf{D}_g(t, \mathbf{r})$ through

$$\mathbf{D}_g = \mathbf{E}_g + 4\pi \int_{-\infty}^{t} \tilde{\mathbf{j}}_g(t', \mathbf{r})dt', \tag{1.13}$$

by use of the above definition and the mass continuity equation(1.12) and by substituting

$$\breve{\phi} = -\phi/4\sqrt{G}, \quad \breve{\mathbf{A}} = -\mathbf{A}/4\sqrt{G},$$
$$\mathbf{E} = -\mathbf{E}_g/4\sqrt{G}, \quad \mathbf{B} = -\mathbf{B}_g/4\sqrt{G}, \quad \mathbf{D} = -\mathbf{D}_g/4\sqrt{G} \tag{1.14}$$

$$\mathbf{j} = \sqrt{G}\rho\mathbf{v}, \quad \hat{\rho} = \sqrt{G}\rho, \quad q = -4\sqrt{G}m, \tag{1.15}$$

the Gravito-Maxwell-Proca equations can be expressed in the standard form,

$$\nabla \times \mathbf{B} = \frac{4\pi}{c}\mathbf{j}_0 + \frac{1}{c}\frac{\partial \mathbf{D}}{\partial t} - \mu_g^2 \breve{\mathbf{A}}, \quad \nabla \times \mathbf{E} = -\frac{1}{c}\frac{\partial \mathbf{B}}{\partial t}, \tag{1.16}$$
$$\nabla \cdot \mathbf{D} = \pi\hat{\rho}_0 - \mu_g^2 \breve{\phi}, \quad \nabla \cdot \mathbf{B} = 0, \tag{1.17}$$

with

$$\mathbf{E} = -\nabla\breve{\phi} - \frac{1}{c}\frac{\partial \breve{\mathbf{A}}}{\partial t}, \quad \nabla \times \breve{\mathbf{A}} = \mathbf{B}; \quad \mathbf{D}(t,\mathbf{r}) = \mathbf{E}(t,\mathbf{r}) + \pi\int_{-\infty}^{t} dt' \mathbf{j}(t',\mathbf{r}); \tag{1.18}$$

and

$$\mathbf{F} = \frac{d\mathbf{p}}{dt} = q\left[\mathbf{E} + \frac{\mathbf{v}}{c} \times \mathbf{B}\right] \tag{1.19}$$

As the graviton has no rest mass the above equations (1.16)--(1.17) become the standard GEM equations (Li, et al., 2009; Li and Liu, 2012)

The field equation (1.16) is in Fourier presentation

$$i(\mathbf{k} \times \mathbf{B}_k)_i = -\frac{i\omega}{c}\varepsilon_{ij}E_{k,j} + \frac{4\pi}{c}j_{0k,i} - \mu_g^2\breve{A}_{k,i},$$
$$\mathbf{k} \times \mathbf{E}_k = \frac{\omega}{c}\mathbf{B}_k,$$

and using the harmonic coordinate condition in Fourier space and the first equation of (1.18),

$$i\mathbf{k} \cdot \breve{\mathbf{A}}_k - i\frac{4}{c}\omega\breve{\phi}_k = 0, \quad \mathbf{E}_k = -i\mathbf{k}\breve{\phi}_k + \frac{i}{c}\omega\breve{\mathbf{A}}_k,$$

One gets

$$\left(\frac{\omega^2}{c^2}k_i\varepsilon_{ij} - \frac{4\omega^2}{4\omega^2 - k^2c^2}k_j\mu_g^2\right)E_{k,j} = -\frac{i4\pi\omega}{c^2}\mathbf{k} \cdot \mathbf{j}_0; \tag{1.20}$$

therefore,

$$\left[\frac{4\omega^2 k}{4\omega^2 - k^2 c^2}\mu_g^2\left(\mathbf{k}/k \cdot \mathbf{e}_{\mathbf{k}}^\sigma\right) - \frac{\omega^2 k}{c^2}\varepsilon^\sigma\right]E_k == \frac{4\pi i}{c^2}\omega \mathbf{k} \cdot \sum_{n\geq 2}\mathbf{j}_k^{T(n)},  \tag{1.21}$$

in which, the nonlinear currents $\mathbf{j}_k^{T(n)} \equiv \mathbf{j}_{\omega,\mathbf{k}}^{T(n)}$ replace the nonlinear response current $\mathbf{j}_0(\omega,\mathbf{k})$ of an external source in (1.20); and

$$\varepsilon^\sigma(\omega,\mathbf{k}) = \varepsilon_{ij}(\omega,\mathbf{k})\frac{k_i e_{\mathbf{k},j}^\sigma}{k}.  \tag{1.22}$$

For the longitudinal field (GE), $e_{\mathbf{k},j}^l = k_j/k$,

$$\left[\frac{4\omega^2}{4\omega^2 - k^2 c^2}\mu_g^2 - \frac{\omega^2}{c^2}\varepsilon_k^l\right]E_k == \frac{4\pi i}{c^2}\omega\frac{\mathbf{k}}{k}\cdot\sum_{n\geq 2}\mathbf{j}_k^{T(n)}  \tag{1.23}$$

with

$$\varepsilon^l(\omega,\mathbf{k}) = \varepsilon_{ij}(\omega,\mathbf{k})k_i k_j/k^2, \quad \mathbf{k}\times\mathbf{E}_{\mathbf{k}} = 0.  \tag{1.24}$$

## II. Kinetic description and linear effects

It is well known that this assumption of short mean free path is invalid for self-gravitating systems with very large scales; one should then define a model of the systems as a collection of dust-like rather than gas-like. Taking into account the presence of massive gravitons (as a dark matter) in the universe, one should treat a two-component self-gravitating system, which is far from a strong gravitational source. The collisionless Boltzmann equations for the distribution function $f_\alpha$ are (Li,1990)

$$\frac{\partial f_\alpha}{\partial t} + \mathbf{v}\cdot\frac{\partial f_\alpha}{\partial \mathbf{r}} + (\mathbf{a}+\mathbf{F})\cdot\frac{\partial f_\alpha}{\partial \mathbf{p}} = 0,  \tag{2.1}$$

where $\mathbf{a}$ is the non-gravitational term(say, centrifugal force), $\mathbf{F}$ is Lorentz-type force (1.19):

$$\mathbf{F} = \frac{d\mathbf{p}}{dt} \approx q_\alpha\left[\mathbf{E} + \frac{\mathbf{v}}{c}\times\mathbf{B}\right],  \tag{2.2}$$

with $q_\alpha = -4\sqrt{G}m_\alpha$; gravitoelectric field $\mathbf{E}$ and gravitomagnetic field $\mathbf{B}$ satisfy (1.16) and(1.17); the density $n_\alpha$ and the mass current density $\mathbf{j}_m$ are connected with the particle distribution through

$$n_a(\mathbf{r},t) = \int f_\alpha(\mathbf{r},\mathbf{v},t) \frac{d\mathbf{p}}{(2\pi)^3}, \quad (\alpha=1,2) \tag{2.3}$$

and

$$\mathbf{j}(\mathbf{r},t) = -\frac{1}{4}\sum_\alpha \int q_\alpha \mathbf{v} f_\alpha(\mathbf{r},\mathbf{v},t) \frac{d\mathbf{p}}{(2\pi)^3}. \tag{2.4}$$

For a large scale gravitational system, not only the distributions of visible matter, including the stars, star clusters, galaxies and medium between the galaxies, are inhomogeneous, but also the dark matter distribution is inhomogenous (Dubovsky *et al.*,2005). As for large-scale dynamics, we regard often these two types of material as "particles" collection, that possess average mass and density over all scales. Therefore, we can generally assume for gravitation effects (Li,1990)

$$\rho_1 = n_0 m_1 \ll \rho_2 = n_0 m_2, \tag{2.5}$$

where $n_0$ is density for the unperturbed state and $\rho_1$ denotes the mass density of bright matter, $\rho_2$ the mass density of dark matter. In the astrophysical accretion disks far from the gravitation source ( black hole or neutron star ), all motion of matter (including rotation) and the nonlinear perturbations caused by matter motion are in the weak fields and slow motion status, so that space-time curvature non-linear effects caused by them are insignificant; in other hand, in the almost flat time-space with the metric(1.6), the kinetic equations describing self-consistently the interactions between GEM fields, gravitons and dust-particles are highly coupled: the distribution functions is not only linked with GEM fields by Eq.(2.1), but also connected to GEM fields through those Maxwell-type equations (1.16) ,(1.17) and the current density Eq.(2.4); in another word, if GEM fields give rise to perturbations of the particle distributions, then through Eq.(2.4), such perturbations in turn induce the perturbations of current density, and consequently, perturb the GEM fields, and so on. Therefore, the response of the medium to the GEM fields is highly nonlinear, which determines the stability, structure, movement and evolution for initial equilibrium system, analogous to the case of laser plasma interactions.

We divide $f_\alpha$ and $\mathbf{E}$ into two parts: unperturbed and perturbed parts,

$$f_\alpha = f_\alpha^R + f_\alpha^T, \quad \mathbf{E} = \mathbf{E}^R + \mathbf{E}^T \tag{2.6}$$

As the $f_\alpha$ is closely coupling with GEM, the perturbed distribution can be expanded in powers

$$f_\alpha^T = \sum_i f^{T(i)}, \tag{2.7}$$

provided that the perturbed field $E^T$ is weak,

$$\bar{W} = \frac{|\mathbf{E}^T|^2}{8\pi n_0 T_0} \ll 1 \tag{2.8}$$

where the superscript $i$ in parentheses denotes the order in $E^T$ and   is the temperature for the unperturbed state. Following Jeans, the instability of interest develops in a homogeneous system; and the unperturbed state is assumed to be in equilibrium. Due to the gravitational acceleration (e.g., rotation), we can choose a uniform background （$\rho_0 = const.$）, so that the unperturbed state is in equilibrium: $\mathbf{a} + \mathbf{F}^R = 0$ (Li, 1990). In this case the equations for the unperturbed state

$$\frac{\partial f_\alpha^R}{\partial t} + \mathbf{v} \cdot \frac{\partial f_\alpha^R}{\partial \mathbf{r}} + (\mathbf{a} + \mathbf{F}^R) \cdot \frac{\partial f_\alpha^R}{\partial \mathbf{p}} = 0 \tag{2.9}$$

are reduced to

$$\frac{\partial f_\alpha^R}{\partial t} = 0 \tag{2.10}$$

with a relevant solution

$$f_{\alpha, p_x}^R \equiv \int f_\alpha^R \frac{dp_y dp_z}{(2\pi)^2} = \frac{(2\pi)^{1/2}}{(m_\alpha v_{T\alpha})} n_0 e^{-\frac{p_x^2}{2m_\alpha^2 v_{T\alpha}^2}}. \tag{2.11}$$

Substituting Eq.(2.6) into Eq.(2.1) yields

$$\frac{\partial f_\alpha^T}{\partial t} + \mathbf{v} \cdot \frac{\partial f_\alpha^T}{\partial \mathbf{r}} + \mathbf{F}^T \cdot \frac{\partial f_\alpha^R}{\partial \mathbf{p}} + \mathbf{F}^T \cdot \frac{\partial f_\alpha^T}{\partial \mathbf{p}} = 0 \ . \tag{2.12}$$

Substituting Eq.(2.7) into Eq.(2.12) and expanding $A = (\mathbf{F}^T, f_\alpha)$ in a Fourier integral

$$A(\mathbf{r}, \mathbf{v}, t) = \int A_k e^{-i\omega t + i\mathbf{k} \cdot \mathbf{r}} dk, \quad A_k \equiv A_{\mathbf{k},\omega}, dk = d\mathbf{k}d\omega,$$

and using the inverse Fourier transformation for product of two functions,

$$(LM)_k = \int L_{k_1} M_{k_2} \delta(k - k_1 - k_2) dk_1 dk_2 \tag{2.13}$$

we get from Eq.(2.12):

$$i(\omega - \mathbf{k} \cdot \mathbf{v}) f_{\alpha,k}^{T(1)} = \int \mathbf{F}_k^T \cdot \frac{\partial f_{\alpha,k_2}^R}{\partial \mathbf{p}} \delta(k - k_1 - k_2) dk_1 dk_2 , \tag{2.14}$$

$$i(\omega - \mathbf{k} \cdot \mathbf{v}) f_{\alpha,k}^{T(2)} = \int \mathbf{F}_{k_1}^T \cdot \frac{\partial f_{\alpha,k_2}^{T(1)}}{\partial \mathbf{p}} \delta(k - k_1 - k_2) dk_1 dk_2 , \tag{2.15}$$

$$i(\omega - \mathbf{k} \cdot \mathbf{v}) f_{\alpha,k}^{T(3)} = \int \mathbf{F}_{k_1}^T \cdot \frac{\partial f_{\alpha,k_2}^{T(2)}}{\partial \mathbf{p}} \delta(k - k_1 - k_2) dk_1 dk_2 ; \tag{2.16}$$

Dividing $\mathbf{j}$ into two parts, $\mathbf{j} = \mathbf{j}^R + \mathbf{j}^T$, and expanding $\mathbf{j}^T$ in powers of $\mathbf{E}^T$

$$\mathbf{j}^{T(i)} = -\frac{1}{4} \sum_\alpha \int q_\alpha \mathbf{v} f_\alpha^{T(i)} \frac{d\mathbf{p}}{(2\pi)^3} ; \tag{2.17}$$

one can obtain the linear current from (2.14) and (2.17)

$$\mathbf{j}_k^{T(1)} = -\frac{1}{4}\sum_\alpha \int \frac{q_\alpha^2 \mathbf{v}\left(\mathbf{E}_k^T + \mathbf{v}\times\left[\frac{\mathbf{k}_1}{\omega_1}\times\mathbf{E}_{k_1}^T\right]\right)\cdot \frac{\partial f_{\alpha,k_2}^R}{\partial \mathbf{p}} dk_2 dk_1}{i(\omega - \mathbf{k}\cdot\mathbf{v} + i\varepsilon)} \delta(k - k_1 - k_2)\frac{d\mathbf{p}}{(2\pi)^3}; \quad (2.18)$$

In which the term $i\varepsilon(\varepsilon \to +0)$ in the denominator of integrated function arises from the Landau rule (Li, 2012). Hence we get from (2.18)

$$j_{k,i}^T = \sigma_{ij}(\omega,\mathbf{k})E_{k,j}^T, \quad (2.19)$$

$$\sigma_{ij}(\omega,\mathbf{k}) = -\frac{1}{4}\sum_\alpha \int \frac{v_i q_\alpha^2\left[\delta_{js}(1-\frac{\mathbf{k}\cdot\mathbf{v}}{\omega}) + \frac{k_s v_j}{\omega}\right]}{i(\omega - \mathbf{k}\cdot\mathbf{v} + i\varepsilon)}\frac{\partial f_\alpha^R}{\partial p_s}\frac{d\mathbf{p}}{(2\pi)^3}, \quad (2.20)$$

here we have taken into account the slow change in $f_\alpha^R$, i.e. $f_{\alpha,k_2}^R \approx f_\alpha^R \delta(k_2)$. By use of the third formula of (1.18) one has

$$D_i(\omega,\mathbf{k}) = \varepsilon_{ij}(\omega,\mathbf{k})E_j(\omega,\mathbf{k}); \quad (2.21)$$

this is the material relation; and $\varepsilon_{ij}$, called the dielectric tensor, is determined as

$$\varepsilon_{ij}(\omega,\mathbf{k}) = \delta_{ij} + \frac{\pi i}{\omega}\sigma_{ij}(\omega,\mathbf{k}). \quad (2.22)$$

Using Eqs.(2.22), and (2.20), one has

$$\varepsilon_k^l = 1 - \sum_\alpha \frac{1}{k^2}\frac{\omega_{p\alpha}^2}{v_{T\alpha}^2}[1 - Z(\frac{\omega}{\sqrt{2}kv_{T\alpha}})], \quad (2.23)$$

where

$$\omega_{p\alpha}^2 \equiv \pi q_\alpha^2 n_0/4m_\alpha = 4\pi G\rho_\alpha, \quad (2.24)$$

$$Z(x) \equiv \int_{-\infty}^\infty \frac{x/\sqrt{\pi}}{x - \xi + i\varepsilon}e^{-\xi^2}d\xi;$$

and $Z(x)$ is the dispersion function (Li, 2012):

$$Z(x) \approx 1 + \frac{1}{2x^2} + \frac{3}{4x^4} - i\sqrt{\pi}xe^{-x^2}, \quad x \gg 1; \quad (2.25)$$

$$Z(x) \approx 2x^2 - i\sqrt{\pi}xe^{-x^2} \approx 2x^2 - i\sqrt{\pi}x, \quad x \ll 1 \quad (2.26)$$

For high-frequency field

$$\omega \gg kv_{T1} \gg kv_{T2}, \quad (2.27)$$

$$\varepsilon_k^l = 1 + \frac{\omega_{p2}^2}{\omega^2} + \frac{\omega_{p1}^2}{\omega^2} + \frac{\omega_{p2}^2}{\omega^2}\frac{3k^2 \mathrm{v}_{T2}^2}{\omega^2} + \frac{\omega_{p1}^2}{\omega^2}\frac{3k^2 \mathrm{v}_{T1}^2}{\omega^4}; \qquad (2.28)$$

for the low-frequency fields, the following conditions are met

$$\omega'/k' \ll \mathrm{v}_{T2}, \quad \omega' \ll \omega_{p1} \ll \omega_{p2}, \qquad (2.29)$$

$$(\varepsilon_{k'}^{2(l)} - 1) \approx -\frac{1}{k^2}\frac{\omega_{p2}^2}{\mathrm{v}_{T2}^2}\left(1 - \frac{\omega^2}{k^2 \mathrm{v}_{T2}^2}\right) = \frac{1}{k^2}\frac{\omega_{p2}^2}{\mathrm{v}_{T2}^2}\left(\frac{\omega^2 - k^2 \mathrm{v}_{T2}^2}{k^2 \mathrm{v}_{T2}^2}\right), \qquad (2.30)$$

with

$$\varepsilon_k^l = 1 + [\varepsilon_k^{1(l)} - 1] + [\varepsilon_k^{2(l)} - 1]. \qquad (2.31)$$

one gets the linear response from (1.23) ($\mathbf{j}_k^{T(n)} = 0$),

$$\left[\frac{4\omega^2}{4\omega^2 - k^2 c^2}\mu_g^2 - \frac{\omega^2}{c^2}\varepsilon_k^l\right] E_k = 0; \qquad (2.32)$$

and the dispersion equation

$$\varepsilon_k^l - \frac{4\mu_g^2 c^2}{4\omega^2 - k^2 c^2} = 0. \qquad (2.33)$$

Taking Eq.(2.28) into consideration, we obtain the dispersion relationship for the longitudinal oscillation with high-frequency:

$$\omega_\ell^2 \simeq \mu_g^2 c^2 (1 + k^2 c^2 / 4\omega_\ell^2) - (\omega_{p2}^2 + \omega_{p1}^2), \quad \omega_\ell \gg kc. \qquad (2.34)$$

It follows that as $\mu_g = 0$ there usually is no steady longitudinal high-frequency mode on the basis of Eq.(2.28), which is consistent with the analysis for the electro-gravitation kinetics(Li,1990; Li *et al.*, 2009 ). Consider only the oscillations with frequency close to the proper frequency of the medium, similar to plasma case (Zakharov, 1984); this means

$$\omega_\ell^2 \simeq \mu_g^2 c^2 (1 + k^2 c^2 / 4\omega_\ell^2) \qquad (2.35)$$

or,

$$\omega_\ell \simeq \mu_g c + \frac{k^2 c}{8\mu_g}, \quad \omega_\ell \gg kc \qquad (2.36)$$

provided that

$$\mu_g^2 c^2 \gg \omega_{p2}^2 = 4\pi G \rho_2, \qquad (2.37)$$

which is the conditions to exist stable longitudinal mode.

In the case of the larger scale in the universe, (2.37) is satisfied with a large reserve : one takes

$\rho_2 \sim 10 (g \cdot cm^{-3})$ in an astrophysical accretion disks ( Li *et al.*, 2009) and then it gets from (2.37),

$$M_{gr} \gg 10^{-52}(g). \tag{2.38}$$

Here we should point out that, due to appear at the proper high frequency oscillation ($\mu_g c$ )of graviton in the linear field equations, it inhibited the oscillation with low frequency for bright material : under the condition of low frequency, (2.32)becomes

$$\left[ \frac{4\omega'^2}{4\omega'^2 - k'^2 c^2} \mu_g^2 - \frac{\omega'^2}{c^2} \varepsilon_{k'}^l \right] E_{k'}^S = 0 ;$$

thus one has $E_{k'}^S = 0$.

## III. High-frequency field equation

Similarly, we get the nonlinear terms (up to the third order) from Eqs.(2.15) and (2.16)

$$f_{\alpha;k}^{(2)} = \int \Sigma_{k,k_1,k_2}^{\alpha} E_{k_1}^{T(+)} E_{k_2}^{T(-)} \delta(k - k_1 - k_2) dk_1 dk_2 \tag{3.1}$$

with

$$\Sigma_{k,k_1,k_2}^{\alpha} = \frac{-q_\alpha^2}{\omega - \mathbf{k} \cdot \mathbf{v} + i\varepsilon} \left\{ \mathbf{e}_{\mathbf{k}_1}^l \cdot \frac{\partial}{\partial \mathbf{p}} \frac{\mathbf{e}_{\mathbf{k}_2}^l \cdot \frac{\partial}{\partial \mathbf{p}}}{\omega_2 - \mathbf{k}_2 \cdot \mathbf{v} + i\varepsilon} + \mathbf{e}_{\mathbf{k}_2}^l \cdot \frac{\partial}{\partial \mathbf{p}} \frac{\mathbf{e}_{\mathbf{k}_1}^l \cdot \frac{\partial}{\partial \mathbf{p}}}{\omega_1 - \mathbf{k}_1 \cdot \mathbf{v} + i\varepsilon} \right\} f_\alpha^R \tag{3.2}$$

and

$$\mathbf{j}_k^{(3)} = \sum_\alpha \int \mathbf{G}_{k,k_1,k_2,k_3}^{\alpha} E_{k_1}^T E_{k_2}^T E_{k_3}^T \delta(k - k_1 - k_2 - k_3) dk_1 dk_2 dk_3 , \tag{3.3}$$

here

$$\mathbf{G}_{k,k_1,k_2,k_3}^{\alpha} = -\frac{1}{4} i q_\alpha^4 \int \mathbf{v} \frac{\mathbf{e}_{\mathbf{k}_1}^l \cdot \frac{\partial}{\partial \mathbf{p}}}{(\omega - \mathbf{k} \cdot \mathbf{v} + i\varepsilon) \left[ (\omega - \omega_1) - (\mathbf{k} - \mathbf{k}_1) \cdot \mathbf{v} + i\varepsilon \right]} \frac{\mathbf{e}_{\mathbf{k}_2}^l \cdot \frac{\partial}{\partial \mathbf{p}}}{} \times$$

$$\frac{\mathbf{e}_{\mathbf{k}_3}^l \cdot \dfrac{\partial}{\partial \mathbf{p}}}{(\omega_3 - \mathbf{k}_3 \cdot \mathbf{v} + i\varepsilon)} f_\alpha^R \frac{d\mathbf{p}}{(2\pi)^3}. \qquad (3.4)$$

To rewrite down field eqation (1.23) for plasmon ($\sigma = l$)

$$\left[\frac{4\omega^2}{4\omega^2 - k^2 c^2}\mu_g^2 - \frac{\omega^2}{c^2}\varepsilon^l\right]E_k^{Tl} = \frac{4\pi i}{c^2}\omega\frac{\mathbf{k}}{k}\cdot\sum_{n\geq 2}\mathbf{j}_k^{T(n)}, \qquad (3.5)$$

here $E_k^{Tl} \equiv E_{\omega,\mathbf{k}}^{Tl}$ is gravito-electric field for the plasmon. If $E_k^T$ in the left-hand side of Eq. (3.5) is high-frequency field, $E_k^T = E_k^{T(+)}$, The three-field product included in the current $\mathbf{j}_k^{(3)}$ can be expressed in terms of high-frequency fields

$$E_{k_1}^T E_{k_2}^T E_{k_3}^T \approx E_{k_1}^{Th} E_{k_2}^{Th} E_{k_3}^{Th};$$

Due to the factor $\left[(\omega - \omega_1) - (\mathbf{k} - \mathbf{k}_1)\cdot\mathbf{v} + i\varepsilon\right]^{-1}$ in the Eq.(3.4), its contribution is important to the $\mathbf{j}_k^{(3)}$ if $E_{k_1}^{Th}$ is the positive high-frequency fields. As a result

$$E_{k_1}^T E_{k_2}^T E_{k_3}^T \approx E_{k_1}^{T(+)}\left[E_{k_2}^{T(+)}E_{k_3}^{T(-)} + E_{k_2}^{T(-)}E_{k_3}^{T(+)}\right],$$

where upper indices "+" and "-" denote the positive and negative frequency parts of high-frequency perturbations, respectively. Hence

$$\mathbf{j}_k^{(3)} = \sum_\alpha\int\mathbf{G}_{k,k_1,k_2,k_3}^\alpha E_{k_1}^{T(+)}\left[E_{k_2}^{T(+)}E_{k_3}^{T(-)} + E_{k_2}^{T(-)}E_{k_3}^{T(+)}\right]\delta(k-k_1-k_2-k_3)dk_1dk_2dk_3 \qquad (3.6)$$

Using the substitutions of that $k_1 \to k_2$ and $k_2 \to k_1$ in the second integral term of Eq.(3.6), it yields

$$\mathbf{j}_k^{(3)} = \sum_\alpha\int\left(\mathbf{G}_{k,k_1,k_2,k_3}^\alpha + \mathbf{G}_{k,k_1,k_3,k_2}^\alpha\right)E_{k_1}^{T(+)}E_{k_2}^{T(+)}E_{k_3}^{T(-)}\delta(k-k_1-k_2-k_3)dk_1dk_2dk_3.$$

Therefore, we get the high-frequency field equation as follows

$$\left(\frac{4\omega^2}{4\omega^2 - k^2 c^2}\mu_g^2 c^2 - \omega^2\varepsilon_k^l\right)E_k^{T(+)} =$$

$$\sum_\alpha\int\tilde{G}_{k,k_1,k_2,k_3}^\alpha E_{k_1}^{T(+)}E_{k_2}^{T(+)}E_{k_3}^{T(-)}\delta(k-k_1-k_2-k_3)dk_1dk_2dk_3, \qquad (3.7)$$

with

$$\tilde{G}_{k,k_1,k_2,k_3}^\alpha = \mathbf{e}_\mathbf{k}^l\cdot\left(\mathbf{G}_{k,k_1,k_2,k_3}^\alpha + \mathbf{G}_{k,k_1,k_3,k_2}^\alpha\right). \qquad (3.8)$$

The symbol $\sum_\alpha$ in those expressions above means adding the contributions of dark matter ($\alpha = 2$) and bright matter ($\alpha = 1$). From Eqs. (3.4) one can see that $f_\alpha^R d\mathbf{p} \sim n_0, q_\alpha \sim m_\alpha$, $G^\alpha \propto m_\alpha$, namely that the matrix elements of the interactions are proportional to the mass of the particles. As the dark matter mass is far larger than the bright one [see Eq. (2.5)], we can neglect the contributions of the bright matter. Therefore, we write $\tilde{G}^\alpha_{k,k_1,k_2,k_3}$ as $\tilde{G}^{\alpha=2}_{k,k_1,k_2,k_3}$

In order to get the field equations in spectrum space for nonlinear interactions up to the third order, we must estimate in detail the integral values of matrix element $\tilde{G}^{\alpha=2}_{k,k_1,k_2,k_3}$. Integrating Eq.(3.4) by parts and by using of Eq.(3.8) and (2.36), we get

$$4\pi i\omega \tilde{G}^{\alpha=2}_{k,k_1,k_2,k_3} \approx 4\omega_{p2}^2 \mathbf{e}^l_\mathbf{k} \cdot \mathbf{e}^l_{\mathbf{k}_1} \frac{1}{n_0} \int \frac{d\mathbf{p}}{(2\pi)^3} \Sigma^{\alpha=2}_{k-k_1,k_2,k_3} \quad . \tag{3.9}$$

Then right side of (3.7) becomes

$$-4\omega_{p2}^2 \int \mathbf{e}^l_\mathbf{k} \cdot (\mathbf{E}^{T(+)}_{k_1} \frac{n^s_{k-k_1}}{n_0}) dk_1 , \tag{3.10}$$

where $n^s_{k'}$ is just the second order of perturbed density (to see (3.1)),

$$n^s_{k'} = \int \frac{f^{(2)}_{2;k'} d\mathbf{p}}{(2\pi)^3} = -q_2^2 \int \frac{1}{\omega' - \mathbf{k}' \cdot \mathbf{v} + i\varepsilon} \left\{ \left(\mathbf{e}^l_{\mathbf{k}_3} \cdot \frac{\partial}{\partial \mathbf{p}}\right) \frac{\mathbf{e}^l_{\mathbf{k}_2} \cdot \frac{\partial}{\partial \mathbf{p}}}{\omega_2 - \mathbf{k}_2 \cdot \mathbf{v}} + \left(\mathbf{e}^l_{\mathbf{k}_2} \cdot \frac{\partial}{\partial \mathbf{p}}\right) \frac{\mathbf{e}^l_{\mathbf{k}_3} \cdot \frac{\partial}{\partial \mathbf{p}}}{\omega_3 - \mathbf{k}_3 \cdot \mathbf{v}} \right\}$$

$$\times E^{T(+)}_{k_2} E^{T(-)}_{k_3} \delta(k' - k_2 - k_3) dk_2 dk_3 \frac{f_2^R d\mathbf{p}}{(2\pi)^3} , \tag{3.11}$$

It describes density fluctuations caused by the scattering of both the positive and negative frequency field on the massive gravitons. Therefore Eq. (3.7) is reduced to

$$\left(\frac{4\omega^2}{4\omega^2 - k^2 c^2} \mu_g^2 c^2 - \omega^2 \varepsilon_k^l\right) E^{T(+)}_k = -4\omega_{p2}^2 \int \mathbf{e}^l_\mathbf{k} \cdot (\mathbf{E}^{T(+)}_{k_1} \frac{n^s_{k-k_1}}{n_0}) dk_1 ; \tag{3.12}$$

note that $E^T_k = E^T_k \mathbf{e}^l_\mathbf{k} \cdot \mathbf{e}^l_\mathbf{k} = \mathbf{E}^T_k \cdot \mathbf{e}^l_\mathbf{k}$, hence

$$\left(\frac{4\omega^2}{4\omega^2 - k^2 c^2} \mu_g^2 c^2 - \omega^2 \varepsilon_k^l\right) \mathbf{E}^{T(+)}_k = -4\omega_{p2}^2 \int (\mathbf{E}^{T(+)}_{k_1} \frac{n^s_{k-k_1}}{n_0}) dk_1 . \tag{3.13}$$

As a result, we obtain from Eq.(3.13)

$$\frac{2i}{\omega_0}\frac{\partial \mathbf{E}(\mathbf{r},t)}{\partial t} + \frac{1}{4}\frac{c^2}{\omega_0^2}\nabla^2 \mathbf{E}(\mathbf{r},t) - 4\frac{\omega_{p2}^2}{\omega_0^2}\frac{n^s(\mathbf{r},t)}{n_0}\mathbf{E}(\mathbf{r},t) = 0, \quad (3.14)$$

where $\mathbf{E}(\mathbf{r},t)$ is the envelope for the high-frequency fields,

$$\mathbf{E}(\mathbf{r},t)e^{-i\omega_0 t} = \int \mathbf{E}_k^{T(+)} e^{-i\omega t + i\mathbf{k}\cdot\mathbf{r}} dk, \quad \omega_0 \equiv \mu_g c \quad (3.15)$$

and because of slow change in $\mathbf{E}(\mathbf{r},t)$, we have neglected the term $\dfrac{\partial^2}{\partial t^2}\mathbf{E}(\mathbf{r},t)$.

## IV. Perturbed density

In order to get the density perturbation in (3.14), we have to estimate the matrix element $\tilde{S}_{k,k_1,k_2}^{2(l)}$, in which $k = (\omega, \mathbf{k})$ belongs to longitudinal field with low-frequency, $\omega_1$ and $\omega_2$ are high-frequency:

$$4\frac{k}{q_2}\tilde{S}_{k,k_1,k_2}^{\alpha=2} = q_2^2 \int \frac{\mathbf{k}\cdot\mathbf{v}+\omega-\omega}{\omega-\mathbf{k}\cdot\mathbf{v}+i\varepsilon}\left\{\left(\mathbf{e}_{\mathbf{k}_1}^l \cdot \frac{\partial}{\partial \mathbf{p}}\right)\frac{\mathbf{e}_{\mathbf{k}_2}^l \cdot \frac{\partial}{\partial \mathbf{p}}}{\omega_2 - \mathbf{k}_2\cdot\mathbf{v}} + \left(\mathbf{e}_{\mathbf{k}_2}^l \cdot \frac{\partial}{\partial \mathbf{p}}\right)\frac{\mathbf{e}_{\mathbf{k}_1}^l \cdot \frac{\partial}{\partial \mathbf{p}}}{\omega_1 - \mathbf{k}_1\cdot\mathbf{v}}\right\}\frac{f_2^R d\mathbf{p}}{(2\pi)^3} \quad (4.1)$$

$$= q_2^2 \omega \int \frac{1}{\omega - \mathbf{k}\cdot\mathbf{v}+i\varepsilon}\left\{\left(\mathbf{e}_{\mathbf{k}_1}^l \cdot \frac{\partial}{\partial \mathbf{p}}\right)\frac{\mathbf{e}_{\mathbf{k}_2}^l \cdot \frac{\partial}{\partial \mathbf{p}}}{\omega_2 - \mathbf{k}_2\cdot\mathbf{v}} + \left(\mathbf{e}_{\mathbf{k}_2}^l \cdot \frac{\partial}{\partial \mathbf{p}}\right)\frac{\mathbf{e}_{\mathbf{k}_1}^l \cdot \frac{\partial}{\partial \mathbf{p}}}{\omega_1 - \mathbf{k}_1\cdot\mathbf{v}}\right\}\frac{f_2^R d\mathbf{p}}{(2\pi)^3}. \quad (4.2)$$

Differentiating with respect to $\mathbf{v}$ in braces in Eq.(4.1) and taking account to delta function in Eq.(3.1) and using (2.23) yields

$$\tilde{S}_{k,k_1,k_2}^{\alpha=2} \approx -\frac{k\omega}{\pi m_2 \omega_{p2}^2} q_2 \left(\varepsilon_k^{2(l)}-1\right)\left(\mathbf{e}_{\mathbf{k}_2}^l \cdot \mathbf{e}_{\mathbf{k}_1}^l\right);$$

(3.11) becomes

$$n_k^s = 4\frac{k^2}{\pi m_2 \omega_{p2}^2}\left(\varepsilon_k^{2(l)}-1\right)\int \mathbf{E}_{k_1}^{T(+)} \cdot \mathbf{E}_{k_2}^{T(-)} \delta(k-k_1-k_2) dk_1 dk_2. \quad (4.3)$$

By the use of (2.30), then, in the coordinate representation, (4.3) becomes

$$v_{T2}^2 \nabla^2 n^s(\mathbf{r},t) = \frac{4/\pi}{k_B T_2}\left(\frac{\partial^2}{\partial t^2} - v_{T2}^2 \nabla^2\right)|\mathbf{E}(\mathbf{r},t)|^2 \quad (4.4)$$

Through the substitution

$$\delta n(\pmb{\xi},\tau)=\frac{n^s(\mathbf{r},t)}{n_0},\ \mathbf{E}(\pmb{\xi},\tau)=\frac{\mathbf{E}(\mathbf{r},t)}{\sqrt{\pi n_0 k_B T_2}},\quad \pmb{\xi}=\frac{4}{c}\omega_{p2}\mathbf{r},\quad \tau=2\frac{\omega_{p2}^2}{\omega_0}t,\quad \beta=\frac{3\omega_{p2}^2}{4\omega_0^2}\frac{c^2}{3v_{T2}^2} \quad (4.5)$$

we can now write Eqs.(3.14) and (4.4) in the form :

$$i\frac{\partial \mathbf{E}(\pmb{\xi},\tau)}{\partial \tau}+\nabla^2 \mathbf{E}(\pmb{\xi},\tau)-\delta n(\pmb{\xi},\tau)\mathbf{E}(\pmb{\xi},\tau)=0, \quad (4.6)$$

$$\left(\beta\frac{\partial^2}{\partial t^2}-\nabla^2\right)|\mathbf{E}(\pmb{\xi},\tau)|^2=\nabla^2 \delta n(\pmb{\xi},\tau), \quad (4.7)$$

where $\nabla$ is defined as the derivative over $\xi$.

## V. Collapse and amplification for the envelope fields

We now restrict ourselves to the state limit, which is relevant in the case of the low-frequency condition (2.29), $\frac{\partial^2}{\partial t^2}\ll \nabla^2$, and so (4.7) is reduced to

$$\delta n(\pmb{\xi},\tau)=-|\mathbf{E}(\pmb{\xi},\tau)|^2 . \quad (5.1)$$

Substituting (5.1) into (4.6) and taking its conjugate, yields

$$i\frac{\partial \mathbf{E}(\pmb{\xi},\tau)}{\partial \tau}-\nabla^2 \mathbf{E}(\pmb{\xi},\tau)-|\mathbf{E}(\pmb{\xi},\tau)|^2 \mathbf{E}(\pmb{\xi},\tau)=0. \quad (5.2)$$

(5.2) is a complex vector field equations and one has so far not finded the analytical solution with more than 2- dimensions. However, we can use the method of field theory to study its properties. Lagrangian density of (5.2) is ( Thornhill and ter Haar,1978)

$$\mathcal{L}=\frac{i}{2}(\mathbf{E}_t^*\cdot \mathbf{E}-\mathbf{E}^*\cdot \mathbf{E}_t)-(\nabla\cdot\mathbf{E})\cdot(\nabla\cdot\mathbf{E}^*)+\frac{1}{2}|\mathbf{E}|^4; \quad (5.3)$$

(5.2) possess following conservation energy

$$\mathcal{E}=\int d\mathbf{r}\left[|\nabla\cdot\mathbf{E}|^2-\frac{1}{2}|\mathbf{E}|^4\right], \quad (5.4)$$

and the conservation "action"

$$N=\int|\mathbf{E}(\pmb{\xi},\tau)|^2 d\pmb{\xi} \quad (5.5)$$

Introducing the scale factor $\lambda(t)$ and taking following transformations

$$\mathbf{r} = \mathbf{r}'/\lambda, \quad \mathbf{E} = \mathbf{E}'\lambda^{3/2}, \tag{5.6}$$

then

$$\nabla = \lambda\nabla', \quad d\mathbf{r} \equiv dxdydz = \lambda^{-3}\,dx'dy'dz' \equiv \lambda^{-3}d\mathbf{r}',$$

(5.4) becomes

$$\mathcal{E}(\lambda) = \int d\mathbf{r}'\left[\alpha\lambda^2\left|\nabla'\times\mathbf{E}'\right|^2 + \lambda^2\left|\nabla'\cdot\mathbf{E}'\right|^2 - \frac{\lambda^3}{2}\left|\mathbf{E}'\right|^4\right]. \tag{5.7}$$

Derivating with respect to $\lambda$ on both sides of the equation (5.7) and putting $\partial\mathcal{E}/\partial\lambda = 0$, one gets the $\lambda_c$ at which $\mathcal{E}$ takes the extreme value,

$$\lambda_c\int d\mathbf{r}'\left|\mathbf{E}'\right|^4 = \frac{4}{3}\int d\mathbf{r}'\left[\alpha\left|\nabla'\times\mathbf{E}'\right|^2 + \left|\nabla'\cdot\mathbf{E}'\right|^2\right]; \tag{5.8}$$

derivating with respect to $\lambda$ on $\partial\mathcal{E}/\partial\lambda$ and using (5.8) one gets

$$\left.\frac{\partial^2\mathrm{E}}{\partial\lambda^2}\right|_{\lambda_c} = -2\int d\mathbf{r}'\left[\alpha\left|\nabla'\times\mathbf{E}'\right|^2 + \left|\nabla'\cdot\mathbf{E}'\right|^2\right] < 0.$$

This means that energy increases with the $\lambda$ and gets maximum at the $\lambda_c$; of course, nonlinear cavity with maximum energy state is not stable: as $\lambda$ goes to $\infty$, then one sees from (5.6), $r = |\mathbf{r}| \to 0$, corresponded to a collapse process - energy ceaselessly decrescent, three dimensional cavity tends to more stable energy state. In other words, 3D nonlinear cavity will be collapsed according to the scale transformation rule (5.6).

Therefore the collapse is actually continuously compressed and "broken" processes: field to be amplified; due to the exclusive "wave pressure", at the same time, there is a density cavity in the increment region of the fields [see (5.1)]. One sees from (5.5) that accompanying the increment region will appear at the decrement area of the fields, where a density peak occurs.

If occasionally gravitational waves in the medium are transversely compressed, then Eq.(5.2) becomes as a nonlinear Schrödinger equation with one-dimension

$$i\frac{\partial}{\partial t}E = -\frac{1}{2}\frac{\partial^2}{\partial\varsigma^2}E - |E|^2\,E.$$

It is shown that (see Li, 2012) there are asymptotically solutions with N – soliton, overtaking each other, and its amplitude and shape does not change when they "collision":

$$E(x,t)|_{t\to\infty} = \sum_{m=1}^{N} u_m, \qquad (5.9)$$

with

$$u_m = -2\sqrt{2}\eta_m \frac{\exp\left[-4i\left(\xi_m^2 - \eta_m^2\right)t - 2i\xi_m x - i\varphi_m\right]}{ch\left[2\eta_m(x - x_{0m}) + 8\xi_m \eta_m t\right]}, (\beta = 1), \qquad (5.10)$$

$\eta_m$ and $\xi_m$ are the parameters associated with soliton amplitude and velocity respectively; $x_{0m}$ and $\varphi_{0m}$ are the center position and phase in solitary wave. In this case, there are electric field peak (soliton) and density cavity.

VI. **Concluding Remarks**

From the above study, we arrive at the following conclusions:

(1) At large distance scales far from a gravitational sourse the linearized Einstein theory is modified accounting for a massive graviton as a dark matter in the universe, this is GEM-Proca equations(1.16)—(1.18). The field equations(1.16)—(1.17)are not invariant for Lorentz gauge; in special the vector potential in the harmonic coordinate conditions has no longer another degree of freedom. This means that the transverse graviton ( $\mathbf{k} \cdot \mathbf{A} = 0$ ) with non-zero mass no contribution to the interaction fields of the system.

(2) On the basis of the GEM-Proca field equations(1.23) in the approximately flat space-time, we investigate the interactions of the wave decay and fusion and wave-particle involving the scattering. It is shown that the dynamic behavior and configuration of the GE fields, the perturbed dark matter density with very low-frequency is determined by Eqs.(4.6)and (4.7).

(3) The resulting envelope equation(5.2)for the high-frequency field is instable with respect to the collapse, leading to enhance of the GE field; In other words, the gravitational waves (GWs) are amplified. As a radiation, i.e. the GWs, enters into this many-particle medium—accretion disk around the neutron star, which is far away from the source of the gravitational radiation, one can estimate the initial wave intensity as (Li et al., 2009)

$$\frac{|\mathbf{E}(\mathbf{r},t)|^2_{\tau=0}}{8\pi n_0 T_0} \equiv \bar{W}_{\tau=0} \sim 10^{-3}, \quad (T_0 \sim T_1);$$

in other hands, the GE fields collapse rapidly and lead to such strong field that $\bar{W} \sim 1$, and in this case the expansion(2.7) is no longer valid. Hance due to self-collapse the perturbed GE is concurrently amplified by a factor of $10^3$; this is the GW cavities describing gravitational shielding and filament effects. And they could appear as the gravitational waves with high energy reaching on Earth. In this case, Weber's results, perhaps, are acceptable.

(4) Meanwhile, the diluting perturbed dark matter density in the local region where the GE enhances and gathering dark matter in field weakening region due to the action conservation [see (5.5)], are in favor of the dark matter clustering and the cavity forming. The results are in good agreement with the observed inhomogeneity properties of dark matter distribution (Dubovsky *et al.*, 2005; Mateo *et al.*, 1993).

(5) Clumps of dark matter, as gravitational seeds, attract to the surrounding bright material and gather them clustering. In this sense, dark matter acts as a compactor of the cluster bright structure.

(6) By the condition to exist stable longitudinal mode[see (2.37)], we can obtain the relevant lower limit for graviton mass: $M_{gr} \geq 10^{-50}(g)$. As a reference, Dubovsky *et al.* (2005) infer that the mass of the graviton can be about $M_{gr} \sim 10^{-52}(g)$, being constrained by the pulsar timing measurements.